**Article type:** **Articles**

# Dynamic Molecular Transformation–Mediated Crystallization Regulation and Interface Passivation in Tin-Based Perovskite Solar Cells

Ying Tang [a,b], Zuhong Zhang [b*], Hongzhuo Wu [b], Zhihao Li [b], Zhenhuang Su [c], Yonggang Yang [d], Hairui Liu [e], Feng Yang [a], Zhe Li [f], Luca Sapienza [g], Jorge Pascual [h], Shengnan Zuo [i], Bo Hou [j*], Antonio Abate [i,k], Meng Li [b,g], Yufang Liu [a,l*]

[a] School of Optoelectronic Engineering, Henan Normal University, Xinxiang, 453007, China

[b] Key Lab for Special Functional Materials of Ministry of Education, State Key Laboratory of Green Chemical Synthesis and Conversion, School of Nanoscience and Materials Engineering, Henan University, Zhengzhou, 450046, China

[c] Shanghai Synchrotron Radiation Facility (SSRF), Shanghai Advanced Research Institute, Shanghai Institute of Applied Physics, Chinese Academy of Sciences, Shanghai, 201204, China.

[d] School of Physics, Henan Normal University, Xinxiang, 453007, China

[e] Inner Mongolia Autonomous University Key Laboratory of Advanced Materials Chemistry and Devices, College of Chemistry and Environmental Science, Inner Mongolia Normal University, Hohhot, 010022, China

[f] School of Engineering and Materials Science (SEMS), Queen Mary University of London, London, E1 4NS, United Kingdom

[g] Department of Engineering, University of Cambridge, Cambridge, CB3 0FA, United Kingdom

[h] Polymat, University of the Basque Country UPV/EHU, Donostia-San Sebastian, 20018, Spain

[i] Helmholtz-Zentrum Berlin für Materialien und Energie GmbH, Hahn-Meitner-Platz 1, Berlin, 14109, Germany

[j] School of Physics and Astronomy, Cardiff University, Cardiff, Wales, CF24 3AA, United Kingdom

[k] Department of Chemical, Materials and Production Engineering, University of Naples Federico II, pzz.le Vincenzo Tecchio 80, Naples, 80125, Italy

[l] Institute of Physics, Henan Academy of Sciences, Zhengzhou, 450000, China

[*]Corresponding Authors: Zuhong Zhang, zhangzuhong@henu.edu.cn; Bo Hou, HouB6@cardiff.ac.uk; Yufang Liu, yf-liu@htu.edu.cn

**Abstract:** Tin-based perovskites are promising candidates for lead-free photovoltaics but suffer from labile $Sn^{2+}$ coordination, rapid crystallization, and severe interfacial defects. Here, we develop a dynamic molecular transformation using 1-acetylpiperazine (PX) to regulate crystallization and interface chemistry simultaneously. During nucleation and growth, the carbonyl group coordinates with $Sn^{2+}$ to stabilize precursor complexes, moderate supersaturation, and enable controlled crystal growth. Upon annealing, PX undergoes in situ molecular transformation, suggested to form amidinium-functionalized species that preferentially occupy surface A-site vacancies, forming a passivation layer and suppressing nonradiative recombination. This coupled bulk-to-interface evolution yields compact, highly crystalline films with improved optoelectronic quality. Devices achieve a power conversion efficiency of 15.18% and retain 94.1% of initial efficiency after 500 h of continuous illumination. This work demonstrates a dynamic molecular engineering approach integrating crystallization control and interface passivation, providing a pathway toward efficient and stable lead-free perovskite photovoltaics.

## 1 Introduction

The escalating global energy crisis and climate change have underscored the urgent demand for efficient, low-cost, and environmentally benign photovoltaic technologies [1, 2]. Perovskite solar cells (PSCs), owing to their superior optoelectronic properties and low fabrication costs, are widely regarded as leading candidates for the next-generation high-performance photovoltaics [3, 4]. However, most state-of-the-art high-efficiency PSCs are predominantly based on lead (Pb)-containing perovskites, where the intrinsic toxicity and sustainability concerns of Pb represent major obstacles to their large-scale deployment [5, 6]. In contrast, tin (Sn), with a similar $ns^2$ $np^2$ electronic configuration and ionic radius to Pb, can form perovskite structures and charge-transport dynamics analogous to Pb-based systems [7], while also offering a close-to-ideal bandgap (~1.4 eV) and strong near-infrared absorption, thereby conferring intrinsic advantages for lead-free technologies [8-11]. Recent advances have boosted their power conversion efficiency (PCE) beyond 17% [12, 13], highlighting their potential in sustainable photovoltaics. Nevertheless, a substantial gap remains compared with the theoretical efficiency limit (>30%) [14], primarily due to severe non-radiative recombination arising from the intrinsic coupling between rapid crystallization and high densities of interfacial defects in tin-based systems, which critically limit both device efficiency and operational stability—critical parameters for the implementation of Sn-based photovoltaic devices in real-life applications.

From the perspective of film formation, tin-based perovskites exhibit a strong propensity for rapid crystallization during deposition. This behavior originates from the stereochemically active $5s^2$ lone pair of $Sn^{2+}$, which induces a more distorted and labile coordination environment and weakens its interaction with Lewis-basic solvents such as dimethyl sulfoxide (DMSO) and dimethylformamide (DMF). As a result, the solvation structure of the precursor undergoes continuous dynamic rearrangement, shortening the nucleation induction time and driving transient supersaturation, thereby triggering rapid high-density nucleation, disordered grain growth, and ultimately yielding to non-compact films [15]. Such structural imperfections not only deteriorate the barrier to moisture and oxygen, accelerating the oxidation of $Sn^{2+}$ to $Sn^{4+}$ and its associated degradation, but also synergistically interact with the volatility and mobility of organic cations during deposition to enrich defects at surfaces and grain boundaries, particularly low-formation-energy formamidinium (FA) vacancies ($V_{FA}$) [16]. These defects substantially promote ion migration and induce local structural reconstruction, slowing exciton relaxation, generating deep-level trap states, and markedly enhancing non-radiative recombination processes, thus severely shortening carrier lifetimes and reducing carrier extraction efficiency, ultimately limiting performance improvements of Sn-based devices [17-19]. To mitigate these issues, a variety of strategies, including ionic doping, surface passivation, and ionic liquid modulation, have been explored to improve crystal quality and suppress surface and grain-boundary defects, yielding notable progress [20-25]. However, these approaches still exhibit inherent limitations in the coupled control of crystallization dynamics and defect chemistry. On one hand, precursor-phase doping typically relies on weak complexation between dopants and $Sn^{2+}$ or organic components. Although such interactions can partially regulate early-stage nucleation, the intrinsically weak coordination of $Sn^{2+}$ and the significant redistribution of dopants during film formation make it difficult to maintain effective control throughout grain growth and defect evolution. On the other hand, surface passivation and post-treatment strategies act primarily on already-formed film interfaces and therefore cannot intervene in the early nucleation and growth process, and as a consequence, they cannot suppress defect formation at the source. In addition, the non-uniform spatial distribution of passivation agents often undermines long-term stability and reproducibility. Collectively, these challenges show the urgent need for a cooperative regulation strategy that can span the entire "nucleation–grain growth" process while simultaneously enabling sustained and precise passivation of interfacial defects, thereby overcoming the core efficiency and stability bottlenecks in tin-based perovskite photovoltaics.

To this end, we propose a dynamic molecular transformation strategy using 1-acetylpiperazine (PX), in which a single molecule exhibits stage-dependent functions during perovskite crystallization and film formation, enabling simultaneous regulation of crystal growth and interfacial defect chemistry. During nucleation and grain growth, the carbonyl (C=O) group of PX coordinates with $Sn^{2+}$ via Lewis acid–base interactions; this coordination stabilizes precursor complexes, moderates supersaturation buildup, and suppresses burst nucleation, thereby promoting more ordered and controllable crystal growth. During annealing, an in situ amidinium transfer process between PX and formamidinium iodide (FAI) is suggested to generate a transient amidinium-functionalized intermediate species (denoted as PXY), which selectively fills $FA^+$ vacancies at grain boundaries and surfaces, constructing a chemically robust, highly directional amidinium-rich passivation network that effectively suppresses shallow trap states and substantially reduces non-radiative recombination. This dual-track regulation, combining nucleation–growth kinetics control with in situ chemical passivation of interfacial defects, enables simultaneous enhancement of device efficiency and operational stability. Benefiting from this strategy, tin-based perovskite films exhibit markedly improved crystallinity, compactness, and interfacial quality; the corresponding devices achieve a PCE of 15.18% and retain 94.1% of their initial efficiency after 500 h of continuous illumination. These results demonstrate the practical potential and generalizability of this molecular engineering approach for efficient and stable lead-free perovskite photovoltaics.

## 2 Experimental section

### 2.1 Materials

The indium tin oxide (ITO) substrate was purchased from Libra Technologies (ST07-15). Poly (3,4-ethylenedioxythiophene)-poly (styrenesulfonate) (PEDOT:PSS) was obtained from Heraeus-Clevios P VP Al 4083. Formamidinium iodide (FAI, >99.99%) and phenethylammonium bromide (PEABr) were purchased from Greatcell Solar Materials. Tin (II) iodide ($SnI_2$, 99.999%), tin (II) fluoride ($SnF_2$, 99.999%), and [6,6]-Phenyl-C61-butyric acid methyl ester (PCBM, 99.9%) were purchased from Libra Technology Corporation. Germanium (II) iodide ($GeI_2$, ≥99.8%) was purchased from Sigma-Aldrich. Ethylenediamine dihydroiodide ($EDAI_2$, 99.5%) was purchased from Xi'an Yuri Solar Company Limited. Bathocuproine (BCP, ＞99.0%) was purchased from TCI. 1-Acetylpiperazine (PX, ≥99%) was purchased from Aladdin Scientific Company. Solvents such as chlorobenzene (CB, 99%, SuperDry, with molecular sieves), *N*,*N*-dimethylformamide (DMF, 99.8%, SuperDry, with molecular sieves), dimethyl sulfoxide (DMSO, 99%, SuperDry, with molecular sieves), and isopropanol (IPA, ≥99.9%, SuperDry, with molecular sieves) were all obtained from J&K Scientific. All solvents and chemicals were used as purchased without any purification.

### 2.2 Perovskite solution preparation

Iodine ($I_2$, 1.52 g, 5.99 mmol) and an excess amount of tin granules (1.19 g, 10.0 mmol) were added to a 20 mL glass vial containing 7.06 mL of a 4:1 (v/v) solvent mixture of DMF and DMSO. The mixture, initially brown in color, was stirred overnight at room temperature, resulting in a clear, bright yellow solution corresponding to a $SnI_2$ concentration of 0.90 M.

A 0.9 M control perovskite precursor solution was prepared by dissolving FAI, $GeI_2$, $SnF_2$, and PEABr in 1mL of the as-prepared $SnI_2$ solution under stirring with an overall molar ratio of FAI:$SnI_2$:$GeI_2$:$SnF_2$:PEABr=1:1:0.05:0.1:0.05. For the preparation of PX-containing perovskite precursors, varying amounts of PX (ranging from 0.25 to 1 μL $mL^{-1}$) were incorporated into the control formulation, with 0.5 μL $mL^{-1}$ identified as the optimal concentration. The precursor solutions were stirred at room temperature for approximately 10 hours to ensure complete dissolution of all components. Finally, the solutions were filtered through a 0.22 μm polytetrafluoroethylene (PTFE) syringe filter before film deposition.

### 2.3 Device fabrication

Indium-doped tin oxide (ITO) glass substrates were sequentially cleaned by ultrasonic treatment in deionized water, acetone, and ethanol for 15 minutes each. After drying, the substrates were exposed to ultraviolet-ozone treatment for 30 minutes to improve surface cleanliness and wettability. A filtered PEDOT:PSS aqueous solution (0.45 μm PTFE filter) was then spin-coated onto the cleaned ITO substrates at 4000 rpm for 50 seconds and subsequently annealed at 140°C for 20 minutes. The resulting ITO/PEDOT:PSS substrates were immediately transferred into a nitrogen-filled glovebox for subsequent processing. Subsequently, the perovskite precursor solution was deposited onto the substrates via spin-coating at 8000 rpm for 60 seconds. During the spin-coating process, 300 μL CB was dropped onto the substrate at 22 seconds to induce film crystallization. The resulting perovskite films were annealed at 100°C for 10 minutes. To further passivate the surface, the $EDAI_2$ solution (0.5 mg $mL^{-1}$ in a 1:1 v/v mixture of IPA and CB) was spin-coated at 4000 rpm for 30 s and subsequently annealed at 100°C for 5 minutes. Next, 100 μL PCBM solution (20 mg $mL^{-1}$ in CB) was deposited on the $EDAI_2$ layer, followed by a brief annealing at 60°C for 1 minute. Finally, 15 nm BCP and 110 nm Ag layers were sequentially deposited via thermal evaporation under high vacuum ($10^{-5}$ Pa). The active area of each fabricated device was 0.0982 $cm^2$.

## 3 Result and discussion

### 3.1 Coordination interactions and defect passivation mechanism of PX in tin-based perovskites

As schematically illustrated in Fig. 1a and Scheme S1, surface $FA^+$ reacts with PX to form $PXY^+$, accompanied by the release of $NH_3$. This transformation was initially evidenced by $^1H$ nuclear magnetic resonance ($^1H$ NMR, Fig. S1) analysis, and the formation of $PXY^+$ is further corroborated by $^1H$ NMR (Fig. S2) and high-resolution electrospray ionization mass spectrometry (ESI-MS, Fig. S3). The molecular electrostatic potential map of PX (Fig. 1b) reveals a pronounced electron-rich region localized around the carbonyl group, enabling coordination with $Sn^{2+}$ through Lewis acid–base interactions [23, 26, 27]. Direct evidence for such coordination is provided by $^1H$ NMR spectra (Fig. 1c), where chemical shift variations at multiple proton sites (f, b, d, a, and e) indicate strong interactions between $SnI_2$ and the electron-donating moieties in PX (e.g., C=O and N-containing sites) [28]. This observation is consistent with Fourier Transform Infrared (FTIR) spectroscopy (Fig. S4), which shows a redshift in the C=O stretching vibration from 1425.8 to 1424.8 $cm^{-1}$. Furthermore, X-ray photoelectron spectroscopy (XPS, Fig. S5) reveals a shift of the Sn 3d peaks toward lower binding energy, indicating a more electron-rich Sn environment that is consistent with coordination between PX and $SnI_2$ [29]. To gain deeper quantitative insight into the defect passivation mechanism, density functional theory (DFT) calculations were performed to evaluate the interactions between PXY and the $FASnI_3$ surface with and without FA vacancies ($V_{FA}$), as well as the corresponding defect formation energies. The optimized structures of vacancy-containing and PXY-coordinated $FASnI_3$ are presented in Fig. 1d. As summarized in Fig. S6, PXY exhibits a binding energy of −2.56 eV on the pristine surface, which becomes more negative (−3.25 eV) in the presence of $V_{FA}$, indicating a thermodynamically more favorable occupation of $V_{FA}$ sites [30]. This preferential binding highlights the strong affinity of PXY to FA-deficient perovskites, and its ability to stabilize the local lattice. In addition, the incorporation of PXY markedly increases the formation energy of $V_{FA}$ (Fig. 1e), showing its pronounced defect passivation capability [31].

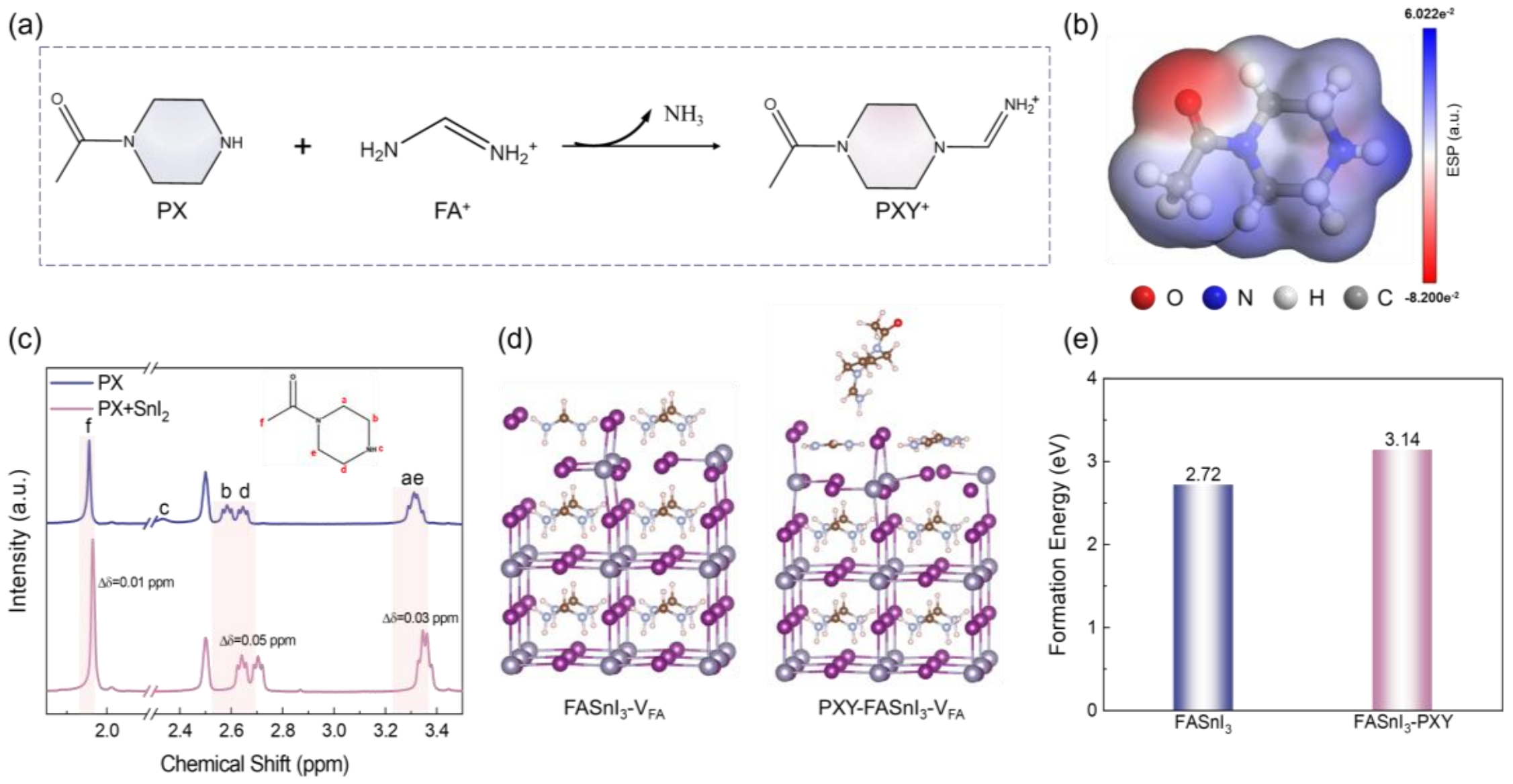

**Fig. 1.** Molecular interaction and passivation mechanism. (a) Schematic illustration of the adduct formation reaction between PX and FAI at the perovskite surface. (b) Electrostatic potential profile of the PX molecule. (c) $^1H$ NMR spectra of PX and PX/$SnI_2$ mixtures in DMSO-$d_6$. (d) DFT-optimized structural models of perovskites with FA vacancies (left), and PXY-coordinated $FASnI_3$ perovskites containing FA vacancies (right). (e) Defect formation

energy of FA vacancies in $FASnI_3$ without and with PXY incorporation.

### 3.2 PX-mediated regulation of crystallization dynamics in $FASnI_3$ films

To unravel how PX regulates film formation, we investigated its influence on crystallization kinetics and film morphology using combined structural and in situ spectroscopic analyses. As illustrated in Fig. 2a, PX coordinates with $Sn^{2+}$ in the precursor solution through Lewis acid–base interactions, forming stabilized intermediates that suppress premature crystallization. In parallel, PX reacts with excess FAI to generate PXY species that preferentially accumulate at grain boundaries and interfaces, where they effectively passivate $V_{FA}$. This dual functionality—modulating crystallization kinetics and mitigating defect formation—facilitates the growth of dense, high-quality $FASnI_3$ films. Morphological characterization corroborates this mechanism: scanning electron microscopy (SEM) images (Figs. 2b and 2c) reveal that the control film contains numerous pinholes, whereas the PX-modified film is significantly more compact and uniform. To directly probe the influence of PX on crystallization dynamics, in situ UV-vis absorption and photoluminescence (PL) measurements were performed during thermal annealing. For the control sample (Fig. 2d), the absorption signal rises sharply at ~4.8 s, indicating rapid nucleation and a transition into a fast crystallization regime. By contrast, the PX-treated film (Fig. 2e) exhibits a delayed onset of characteristic absorption at ~7.3 s and a more gradual spectral evolution, confirming that PX–$Sn^{2+}$ coordination moderates the reaction kinetics and enables more controlled grain growth [25, 27, 32]. In situ PL measurements further reinforce this conclusion. The control sample (Fig. 2f) exhibits an abrupt rise in PL intensity at the early stage of annealing, reflecting rapid nucleation and localized radiative recombination within small or disordered crystallites [33, 34]. This transient enhancement is followed by pronounced fluctuations and partial quenching, suggesting defect accumulation and incomplete grain coalescence [35, 36]. In sharp contrast, the PX-modified film (Fig. 2g) displays a delayed and gradually increasing PL response, accompanied by fewer fluctuations and earlier stabilization. This evolution reflects a more gradual and coherent crystallization process. This monotonic PL evolution further suggests suppressed trap formation, consistent with the defect-passivation role of PXY. Together, these results demonstrate that PX simultaneously regulates nucleation and growth kinetics and enables in situ defect passivation via its PXY derivative, thereby yielding compact, defect-suppressed perovskite films with superior morphological and potentially optoelectronic quality.

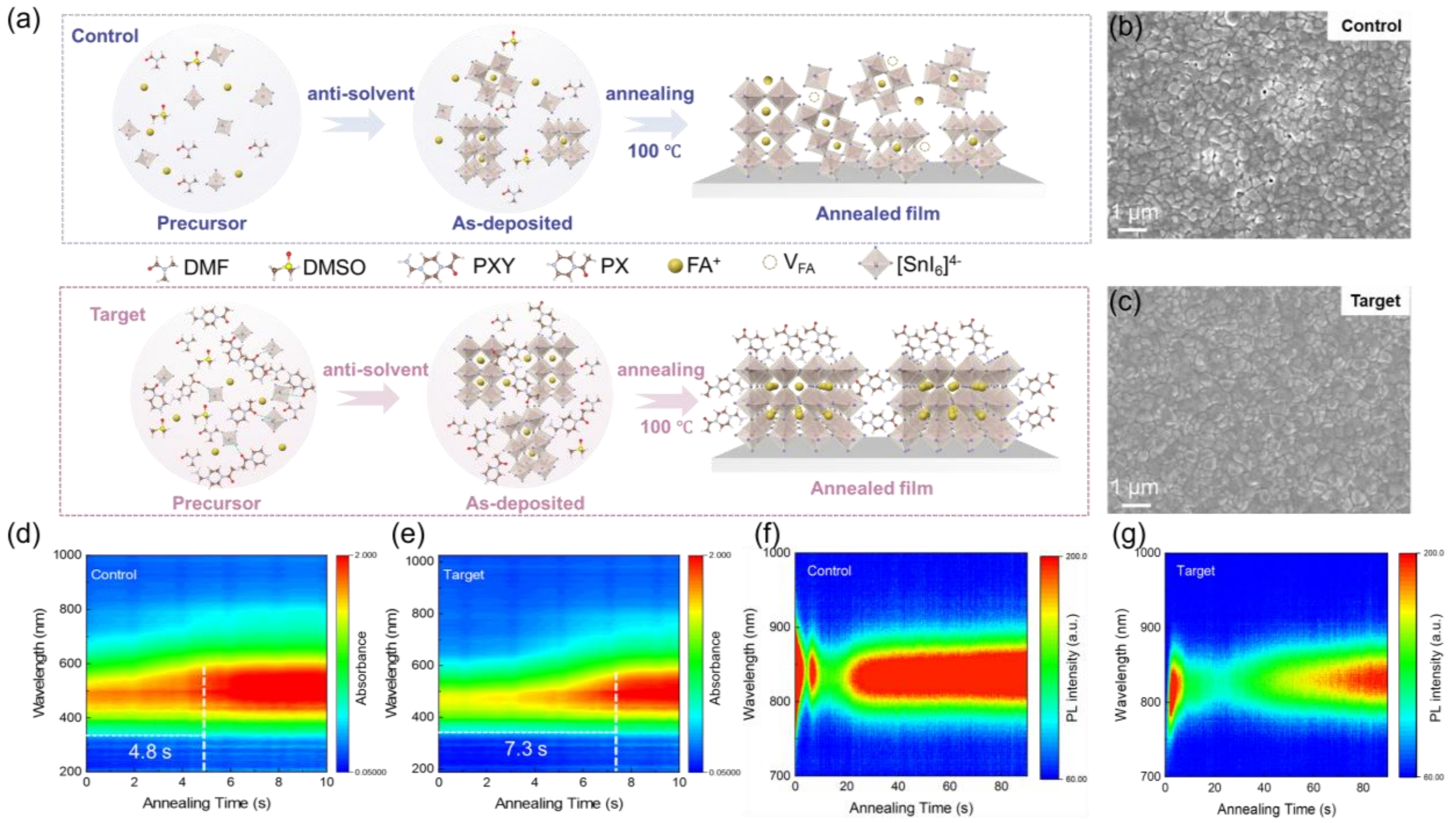

**Fig. 2.** Regulatory mechanism of crystallization kinetics. (a) Proposed schematic illustration of PX-mediated regulation of crystallization and defect passivation in tin-based perovskites. Top-view SEM images of the control (b) and target (c) perovskite films. In-situ UV-vis absorption evolution of the control (d) and target (e) films during the annealing process. In-situ PL evolution of the control (f) and target (g) films during the annealing process.

### 3.3 Defect passivation-mediated charge-carrier dynamics enhancement in PX-modified perovskites

To elucidate the defect passivation mechanism of PX and its PXY derivative and their impact on charge-carrier dynamics, we systematically investigated the structural, optical, and electrical properties of the perovskite films and devices. Structurally, X-ray diffraction (XRD) patterns (Fig. 3a) reveal that PX-treated films exhibit a significantly enhanced (100) diffraction peak, indicating improved crystallinity [37], consistent with the more compact and uniform morphology observed in SEM images. This conclusion is further corroborated by grazing-incidence wide-angle X-ray scattering (GIWAXS), where intensified diffraction signals at different incident angles of 0.1°, 0.4°, and 0.8° (Figs. S7 and S8) confirm that the crystalline quality is significantly improved in both the near-surface and bulk regions [38]. The structural improvement is reflected in the optical absorption characteristics. The Urbach energy ($E_u$) derived from the UV-vis absorption was employed to quantify the degree of structural disorder [39]. As shown in Fig. 3b, $E_u$ decreased markedly from 193.05 meV for the control film to 125.16 meV for the PX-modified film, indicating a substantial reduction in the density of sub-bandgap tail states and an enhanced structural quality of the perovskite film [38, 40-42]. This improvement is further validated by spatially resolved PL mapping (Figs. 3c and 3d), which reveals stronger and more uniform luminescence intensity after PX incorporation, reflecting suppressed non-radiative recombination and more uniform optoelectronic properties [43].

To probe defect passivation at the device level, trap density of states (tDOS) was analyzed via admittance spectroscopy (Fig. S9). PX-modified devices exhibited a pronounced reduction in trap states, particularly in the energy range of 0.275–0.350 eV, which is typically associated with grain-boundary and interfacial defects [44], consistent with the effective passivation of FA-vacancy-related defects by PXY species [45]. Drive-level capacitance profiling (DLCP, Fig. 3e) further corroborates this finding, revealing a substantial decrease in interfacial trap density at the ETL/perovskite interface, which is critical for suppressing interfacial non-radiative recombination [46-48]. Finally, space-charge-limited current (SCLC) measurements (Fig. 3f) provide a quantitative estimate of bulk defect density: the PX-treated devices exhibit a lower defect density ($5.04\times10^{16}$ $cm^{-3}$) compared to the control ($6.05\times10^{16}$ $cm^{-3}$), further confirming that PXY concurrently mitigates both bulk and interfacial defects.

Benefiting from the reduced defect density and improved crystallinity, PX-modified tin halide perovskite solar cells (THPSCs) exhibit substantially improved charge carrier dynamics. Transient photovoltage (TPV, Fig. S10) measurements show that the carrier lifetime increases from 8.50 μs to 22.94 μs, while transient photocurrent (TPC, Fig. S11) measurements revealed that the charge-extraction time is shortened from 1.00 μs to 0.56 μs, evidencing suppressed recombination and more efficient charge extraction [49]. $V_{OC}$-light intensity dependence curves (Fig. S12) further exhibited a reduced slope (1.15 KT/q vs. 1.76 KT/q), indicative of suppressed trap-assisted recombination [50, 51]. Furthermore, complementary electrical characterizations also support this improvement: dark *J-V* curves (Fig. S13) displayed lower leakage currents, reflecting suppressed shunting pathways and reduced trap-assisted recombination [52]. Electrochemical impedance spectroscopy (EIS) Nyquist plots (Fig. S14) revealed a larger recombination resistance, increasing from 2.8 kΩ for the control device to 3.8 kΩ for the target device, indicating lower trap-assisted recombination at both the bulk and interface levels [26, 53]. Moreover, Mott–Schottky analysis (Fig. S15) revealed an increase in built-in potential (0.78 V vs. 0.64 V), indicating a more efficient charge separation, which contributes to an enhancement of open-circuit voltage ($V_{OC}$) [54-57]. Taken together, these results demonstrate that PX improves the crystalline quality and structural order of tin-based perovskite films during the crystallization process, while its PXY derivative allows in situ defect passivation at grain boundaries and interfaces. This synergistic regulation of bulk and interfacial defect landscapes leads to suppressed non-radiative recombination, facilitates charge separation and carrier transport, thereby underpinning the improved optoelectronic performance of THPSCs.

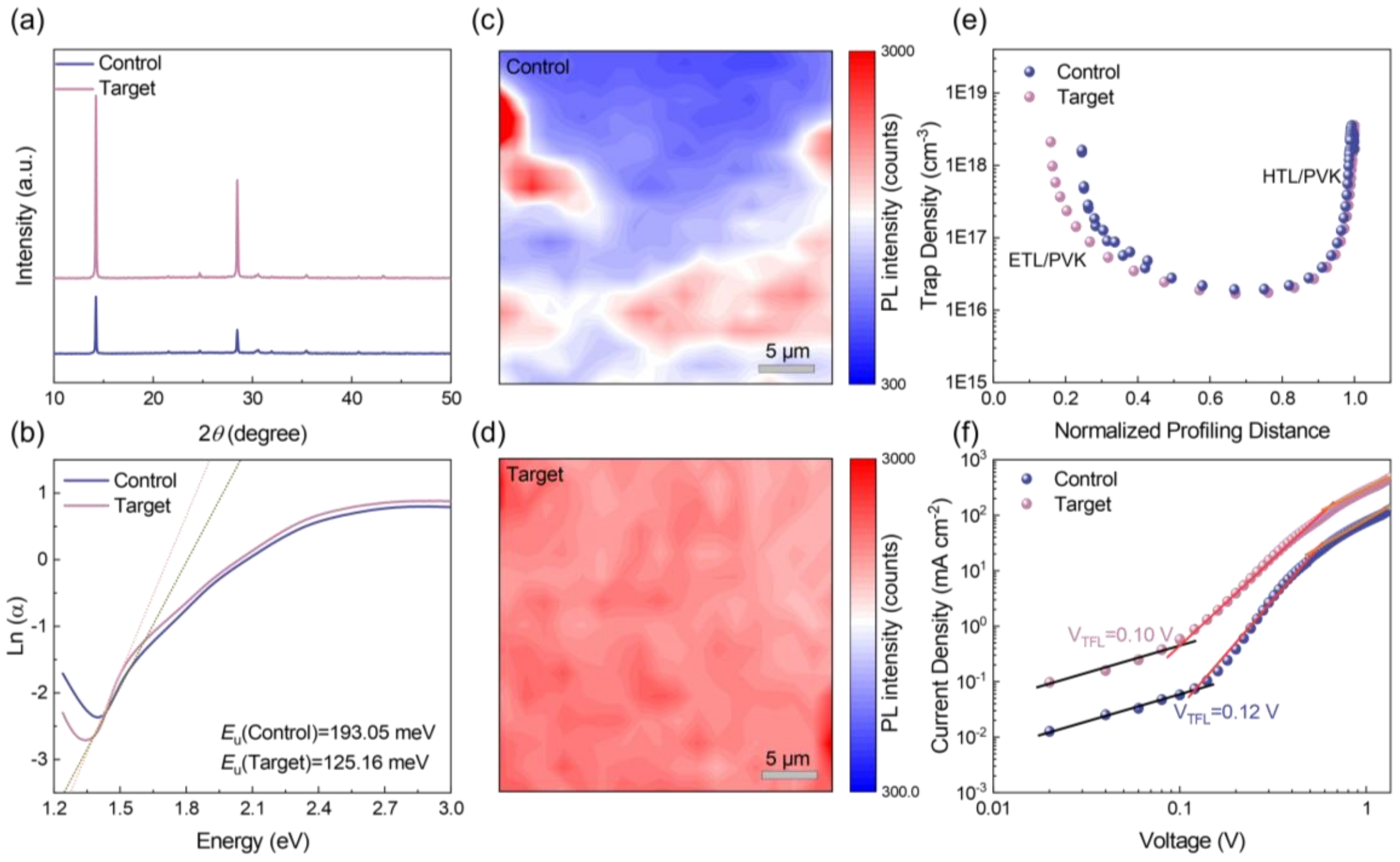

**Fig. 3.** PX-induced film quality enhancement and defect mitigation. (a) XRD patterns of the control and target perovskite films. (b) $E_u$ plots of the control and target perovskite films. PL mapping of the control (c) and target (d) perovskite films deposited on glass substrates. (e) DLCP profiles of control and target devices. (f) *J-V* curves of the electron-only devices.

### 3.4 Photovoltaic performance and stability of THPSCs

To evaluate the impact of PX on the photovoltaic performance and stability of devices, THPSCs with the structure of ITO/PEDOT:PSS/PVK (with or without PX) /$EDAI_2$/PCBM/BCP/Ag were fabricated. Cross-sectional SEM images (Fig. 4a) show that PX incorporation leads to enlarged grains and more vertically aligned crystal domains, reflecting improved film quality. Based on the optimization summarized in Fig. S16 and Table S1, an optimal PX concentration of 0.50 μL $mL^{-1}$ was selected for subsequent device fabrication. As shown in Fig. 4b, the control device delivered a PCE of 13.74%, with a short-circuit current density ($J_{SC}$) of 21.47 mA $cm^{-2}$, an open-circuit voltage ($V_{OC}$) of 0.90 V, and a fill factor (FF) of 71.1%. Upon the introduction of PX, the champion target devices achieved a PCE of 15.18%, with a slightly higher $J_{SC}$ (21.60 mA $cm^{-2}$), an increased $V_{OC}$ (0.92 V), and an enhanced FF (76.4%). The detailed photovoltaic parameters of the control and target devices were presented in Table S2. Statistical analyses of photovoltaic parameters obtained from 12 individual devices (Fig. S17) corroborate the high reproducibility of the improvements. The integrated current densities from the external quantum efficiency (EQE) spectra in Fig. S18 (20.40 and 21.17 mA $cm^{-2}$ for the control and target devices, respectively) are in close agreement with *J-V* results, validating measurement reliability. To further elucidate the origin of efficiency enhancement, a detailed FF loss analysis was conducted (Fig. 4c). Relative to the Shockley–Queisser (SQ) limit, PX-modified devices exhibit a notable reduction in non-radiative recombination losses as well as minimized charge transport losses. The first is attributed to effective passivation of bulk and interfacial defects, while the mitigation of transport-related losses arises from improved carrier extraction and transport efficiency [58].

Beyond performance enhancement, PX also significantly improves material and device stability. XPS analysis (Fig. S19) demonstrates that the initial $Sn^{4+}$ fraction in freshly prepared films is reduced from 34.2% (control) to 17.8% (PX-modified), indicating suppressed $Sn^{2+}$ oxidation. This improvement is consistent with the coordination between the C=O group in PX and $Sn^{2+}$, which likely strengthens the $Sn^{2+}$ coordination environment. To further validate this effect, we compared the appearance of $SnI_2$·DMSO and $SnI_2$·DMSO + PX solutions after one week in an $N_2$ glovebox. The $SnI_2$·DMSO solution gradually changed from pale yellow to orange-yellow, indicative of $Sn^{2+}$ oxidation, whereas the PX-containing solution retained its original color (Fig. S20), confirming the stabilization effect of PX on $Sn^{2+}$. To probe the air-oxidation resistance of the films, as-prepared perovskite layers were taken out of the glovebox and exposed to ambient air for 5 h before a second XPS measurement. Under these conditions, the $Sn^{4+}$ fraction in PX-modified films increased only slightly from 17.8% to 24.4%, whereas that in control films rose sharply from 34.2% to 55.6% (Fig. S21), showing the oxidation resistance imparted by PX. Consistently, morphology comparison (Fig. S22) reveals that PX incorporation significantly suppresses film degradation. This enhanced stability arises from the combined effects of chemical stabilization and morphological improvement, including higher crystallinity and reduced grain-boundary density, which together hinder oxygen/moisture ingress and retard $Sn^{2+}$ oxidation [59, 60]. In line with these material-level advantages, PX-treated devices also exhibit improved operational stability. As shown in Fig. S23, the optimized device treated with PX yielded a stabilized efficiency of 14.9% at the maximum power point (MPP) with a bias voltage of 0.76 V, outperforming the control device, which demonstrated a lower efficiency of 12.0% at 0.70 V. Finally, we record the shelf stability and operational stability of the unencapsulated devices by performing MPP tracking in the $N_2$ atmosphere at room temperature, respectively. Long-term shelf stability tests (Fig. 4d) reveal that PX-modified devices retained 94.2% of their initial PCE after 2000 h in an inert atmosphere, compared to only 50.3% for the control ones. Moreover, during continuous illumination (Fig. 4e), PX-treated devices maintained over 94% of their initial efficiency after 500 h, whereas the control degraded to ~50%. These results highlight the synergistic role of PX in regulating crystallization and suppressing defect-mediated degradation pathways, thereby substantially enhancing both performance and long-term durability of THPSCs.

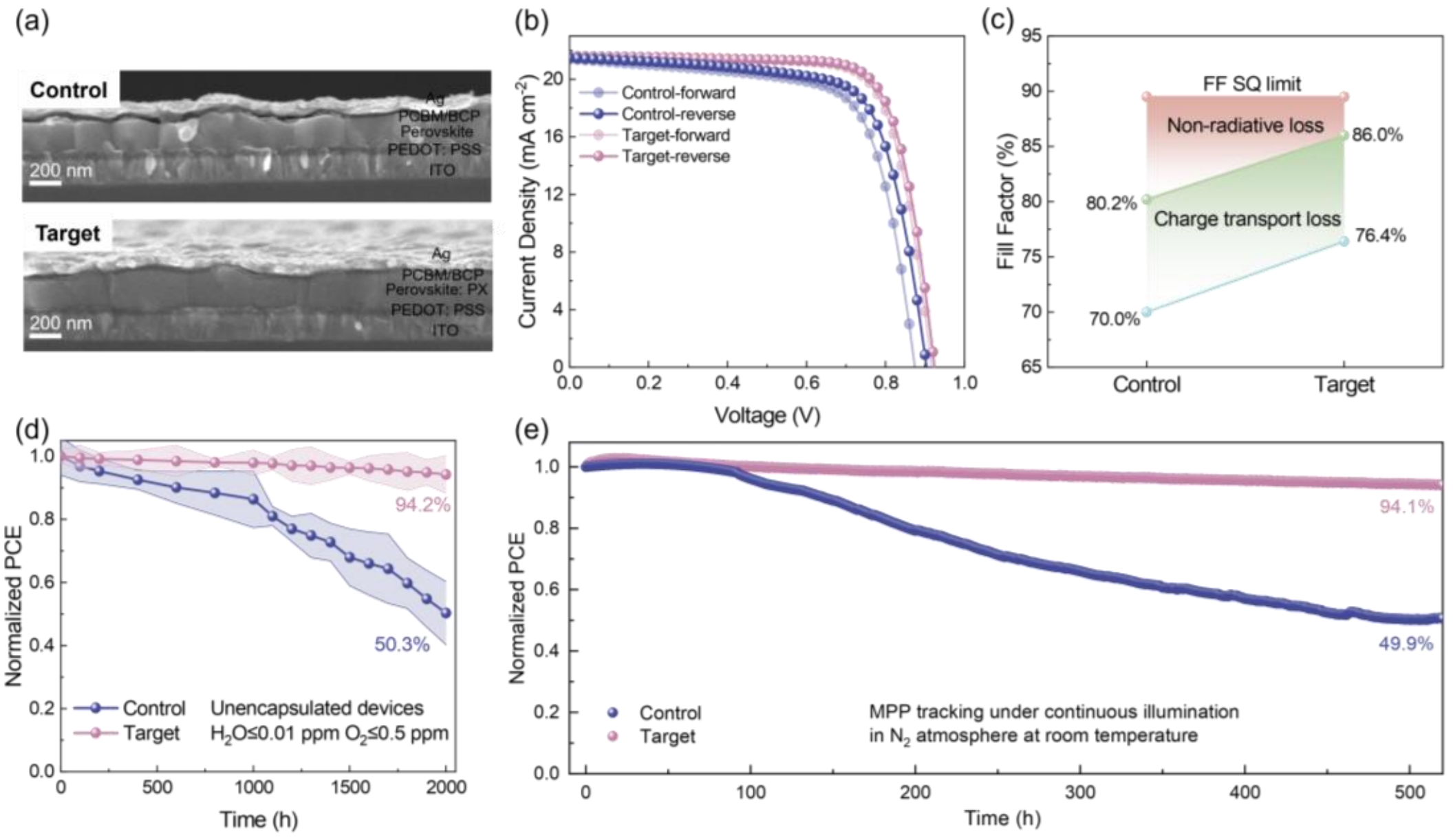

**Fig. 4.** Photovoltaic performance and stability of devices. (a) Cross-sectional SEM images of control and target devices. (b) *J-V* curves of the champion control and target devices. (c) FF SQ limit of the corresponding devices. (d) Shelf stability of the unencapsulated devices under dark conditions in $N_2$ atmosphere. The initial PCEs of the control and target devices were 13.56% and 15.12%, respectively. The solid lines with symbols represent the mean normalized PCE of six individual devices (n=6), and the shaded regions denote ±1 standard deviation (s.d.) across the devices. (e) Long-term operational stability of unencapsulated devices under continuous white LED illumination (100 mW $cm^{-2}$) with MPP tracking in an $N_2$ atmosphere at room temperature. The initial PCEs of the control and target devices were 13.74% and 15.18%, respectively.

## 4 Conclusion

In summary, we have developed a dynamic molecular transformation strategy to address the coupled challenges of rapid crystallization and interfacial defect formation that limit the performance of tin-based perovskites. By incorporating PX into the precursor, we introduce an organic species whose carbonyl group coordinates with $Sn^{2+}$. This coordination is consistent with the delayed onset and smoother evolution of crystallization observed in in situ UV-vis and PL measurements, leading to $FASnI_3$ films with more controlled grain growth, enhanced crystallinity, and increased compactness. During thermal annealing, PX is suggested to undergo an in situ molecular transformation with formamidinium iodide, leading to species that preferentially interact with surface A-site vacancies and contribute to interfacial defect passivation. As a consequence, the resulting THPSCs achieve a champion PCE of 15.18% and retain 94.1% of their initial efficiency after 500 h of continuous illumination. These findings identify amidinium-transfer–driven, stage-adaptive additives as an effective approach to simultaneously regulate crystallization kinetics and passivate interfacial A-site vacancies, providing a promising molecular and interface-engineering basis for efficient and stable lead-free perovskite photovoltaics.

**Conflict of interest**

The authors declare that they have no conflict of interest.

**Acknowledgments**

The authors gratefully acknowledge the technical support from the staff of Henan Normal University and Henan University. Synchrotron measurements were carried out at beamline BL14B1 of the Shanghai Synchrotron Radiation Facility (SSRF), and the authors thank the beamline scientists for providing beamtime and experimental assistance. Y. Liu acknowledges financial support from the Zhongyuan Scholar of Henan Province (No. 224000510007). M. Li acknowledges support from the National Natural Science Foundation of China General Fund (No. 52472199), and the Outstanding Youth Fund of the Natural Science Foundation of Henan Province (No. 242300421069). B. Hou acknowledges financial support from EPSRC SWIMS (EP/V039717/1) and Leverhulme Trust (RPG-2022-263).

**Author contributions**

Ying Tang, Zuhong Zhang, Bo Hou, Yufang Liu, and Meng Li conceived and designed the project. Ying Tang led the fabrication and optimization of the perovskite devices and related thin films, carried out the main electrical and photovoltaic characterizations, and analyzed the data. Zuhong Zhang, Hongzhuo Wu, Zhihao Li, Yonggang Yang, and Hairui Liu assisted with device fabrication, process optimization, and optoelectronic measurements, including *J*–*V*,

EQE, EIS, and Mott–Schottky tests. Zhenhuang Su carried out grazing-incidence wide-angle X-ray scattering (GIWAXS) measurements at the Shanghai Synchrotron Radiation Facility and contributed to the analysis of the structural evolution. Feng Yang conducted SEM characterizations and contributed to correlating morphology with device performance. Ying Tang wrote the original draft. Zhe Li, Luca Sapienza, Jorge Pascual, Shengnan Zuo, Antonio Abate, Yufang Liu, and Meng Li contributed to data interpretation and critical revision of the manuscript. Bo Hou, Yufang Liu, and Meng Li supervised the overall project and acquired funding. All authors discussed the results and contributed to the final version of the manuscript.

**Appendix A. Supplementary material**

Supplementary data to this article can be found on the journal website.

**References**

[1] Omer AM. Energy, environment and sustainable development. Renew Sustain Energy Rev 2008; 12: 2265-2300.

[2] Al-Shetwi AQ. Sustainable development of renewable energy integrated power sector: trends, environmental impacts, and recent challenges. Sci Total Environ 2022; 822: 153645.

[3] Liu T, Ren Z, Liu Y, et al. Efficient perovskite solar modules enabled by a UV-stable and high-conductivity hole transport material. Sci Adv 2025; 11: eadu3493.

[4] Han Y, Fu J, Ren Z, et al. Inorganic perovskite/organic tandem solar cells with 25.1% certified efficiency via bottom contact modulation. Nat Energy 2025; 10: 513-525.

[5] Schileo G, Grancini G. Lead or no lead? Availability, toxicity, sustainability and environmental impact of lead-free perovskite solar cells. J Mater Chem C 2021; 9: 67-76.

[6] Zhang H, Lee J-W, Nasti G, et al. Lead immobilization for environmentally sustainable perovskite solar cells. Nature 2023; 617: 687-695.

[7] Wu T, Liu X, Luo X, et al. Lead-free tin perovskite solar cells. Joule 2021; 5: 863-886.

[8] Liu X, Wu T, Zhang C, et al. Interface energy-level management toward efficient tin perovskite solar cells with hole-transport-layer-free structure. Adv Funct Mater 2021; 31: 2106560.

[9] Imran T, Rauf S, Raza H, et al. Methylammonium and bromide-free tin-based low bandgap perovskite solar cells. Adv Energy Mater 2022; 12: 2200305.

[10] Liao W, Zhao D, Yu Y, et al. Lead-free inverted planar formamidinium tin triiodide perovskite solar cells achieving power conversion efficiencies up to 6.22%. Adv Mater 2016; 28: 9333-9340.

[11] Hao F, Stoumpos CC, Cao DH, et al. Lead-free solid-state organic–inorganic halide perovskite solar cells. Nat Photonics 2014; 8: 489-494.

[12] He D, Chen P, Steele JA, et al. Homogeneous 2D/3D heterostructured tin halide perovskite photovoltaics. Nat Nanotechnol 2025; 20: 779-786.

[13] Li T, Luo X, Wang P, et al. Tin-based perovskite solar cells with a homogeneous buried interface. Nature 2025; 648: 84-90.

[14] Xu L, Feng X, Jia W, et al. Recent advances and challenges of inverted lead-free tin-based perovskite solar cells. Energy Environ Sci 2021; 14: 4292-4317.
[15] Zhang Y, Li C, Zhao H, et al. Synchronized crystallization in tin-lead perovskite solar cells. Nat Commun 2024; 15: 6887.
[16] Li N, Tao S, Chen Y, et al. Cation and anion immobilization through chemical bonding enhancement with fluorides for stable halide perovskite solar cells. Nat Energy 2019; 4: 408-415.
[17] Jiang Q, Tong J, Xian Y, et al. Surface reaction for efficient and stable inverted perovskite solar cells. Nature 2022; 611: 278-283.
[18] Liu C, Yang Y, Chen H, et al. Bimolecularly passivated interface enables efficient and stable inverted perovskite solar cells. Science 2023; 382: 810-815.
[19] Ghosh A, Kumar S, Sarkar P. Point defect-mediated hot carrier relaxation dynamics of lead-free $FASnI_3$ perovskites. Nanoscale 2024; 16: 4737-4744.
[20] Qi J, Zhang B, Ma Y, et al. Interface field-effect passivation enabled by selectively extruded cations for printable mesoscopic perovskite solar cells with improved performance. Adv Mater 2025; 37: 2506114.
[21] Fan X, Chen J, Wang J, et al. Efficient rigid and flexible perovskite solar cells using strongly adsorbed molecules for lattice repair and grain boundary mitigation. Energy Environ Sci 2025; 18: 2905-2917.
[22] Guo Q, Zhang J, Liu J, et al. Promoting charge-carriers dynamics by relaxed lattice strain in A-site-doped halide perovskite for photocatalytic $H_2$ evolution. Angew Chem Int Ed 2025; 64: e202419082.
[23] Ji H, Zhang B, Zhao Q, et al. Organic molecule vapor-assisted passivation for efficient and stable perovskite solar cells. Adv Energy Mater 2025; 15: 2405441.
[24] Wang M, Zhao Y, Jiang X, et al. Rational selection of the polymeric structure for interface engineering of perovskite solar cells. Joule 2022; 6: 1032-1048.
[25] Xu X, Li S, Shan C, et al. Unraveling the interfacial homogeneity and bulk crystallization for efficient and stable perovskite solar cells via ionic liquids. Energy Environ Sci 2025; 18: 3407-3417.
[26] Zhang W, Cai Y, Liu H, et al. Organic-free and lead-free perovskite solar cells with efficiency over 11%. Adv Energy Mater 2022; 12: 2202491.
[27] Li C, Zhu M, Zhang S, et al. Competitive-coordination-induced crystallization regulation for efficient and stable Sn–Pb perovskite solar cells. Angew Chem Int Ed 2025; 64: e202501188.
[28] Duan C, Zhang K, Peng Z, et al. Durable all-inorganic perovskite tandem photovoltaics. Nature 2025; 637: 1111-1117.
[29] Zhou X, Peng W, Liu Z, et al. Additive engineering with 2,8-dibromo-dibenzothiophene-s,s-dioxide enabled tin-based perovskite solar cells with 14.98% power conversion efficiency. Energy Environ Sci 2024; 17: 2837-2844.
[30] Cai Q, Tan Q, He J, et al. Enhancing electron transport for efficiency-recorded HTL-free inverted perovskite solar cells by molecular complementary passivation. Joule 2025; 9: 101880.
[31] Zhang Z, Li M, Li R, et al. Suppressing ion migration by synergistic engineering of anion and cation toward high-performance inverted perovskite solar cells and modules. Adv Mater 2024; 36: 2313860.
[32] Yu B, Wang K, Sun Y, et al. Minimizing buried interface energy losses with post-assembled chelating molecular bridges for high-performance and stable inverted perovskite solar cells. Adv Mater 2025; 37: 2500708.
[33] Xu Z, Guo Z, Li H, et al. Efficient and stable inverted MA/Br-free 2D/3D perovskite solar cells enabled by α-to-δ phase transition inhibition and crystallization modulation. Energy Environ Sci 2025; 18: 1354-1365.

[34] Zhang G, Lv Y, Su Z, et al. Three birds with one stone: triple defect passivation of tris(2,2,2-trifluoroethyl) phosphate enables 25.69% efficient inverted perovskite solar cells. Angew Chem Int Ed 2025; 64: e202502244.
[35] Shao C, Ma J, Niu G, et al. Strain release via glass transition temperature regulation for efficient and stable perovskite solar cells. Adv Mater 2025; 37: 2417150.
[36] Wang S, Yang T, Yang Y, et al. In situ self-elimination of defects via controlled perovskite crystallization dynamics for high-performance solar cells. Adv Mater 2023; 35: 2305314.
[37] Xie G, Li H, Fang J, et al. Crystallization thermodynamics regulation of 1.85 eV wide-bandgap perovskite for efficient and stable perovskite-organic tandem photovoltaics. Angew Chem Int Ed 2025; 64: e202501764.
[38] Jung S, Jang Y, Jung H, et al. Radical scavenger-driven oxidation prevention and structural stabilization for efficient and stable tin-based perovskite solar cells. Energy Environ Sci 2025; 18: 6076-6084.
[39] Kim G, Min H, Lee KS, et al. Impact of strain relaxation on performance of α-formamidinium lead iodide perovskite solar cells. Science 2020; 370: 108-112.
[40] Jiang X, Wang F, Wei Q, et al. Ultra-high open-circuit voltage of tin perovskite solar cells via an electron transporting layer design. Nat Commun 2020; 11: 1245.
[41] Xu P, Chen X, Hui W, et al. Retarding the growth kinetics of chemical bath deposited nickel oxide films for efficient inverted perovskite solar cells and minimodules. Adv Mater 2025; 37: 2505087.
[42] Lv Y, Yang C, Su Z, et al. Molecular integration of Lewis bases for efficient and stable inverted perovskite solar cells. Energy Environ Sci 2025; 18: 6237-6247.
[43] Qu Z, Zhao Y, Ma F, et al. Enhanced charge carrier transport and defects mitigation of passivation layer for efficient perovskite solar cells. Nat Commun 2024; 15: 8620.
[44] Hu Q, Chen W, Yang W, et al. Improving efficiency and stability of perovskite solar cells enabled by a near-infrared-absorbing moisture barrier. Joule 2020; 4: 1575-1593.
[45] Lou Q, Xu X, Xu Z, et al. A two-in-one additive strategy for poly(3-hexylthiophene)-based perovskite solar cells with improved efficiency and stability. Cell Rep Phys Sci 2025; 6: 102515.
[46] Chen X, Chen C, Su Z, et al. Adhesively bridging SAM molecules and perovskites for highly efficient photovoltaics. Adv Funct Mater 2025; 35: 2415004.
[47] Li Y, Xia Y, Chen C, et al. Surficial homogenic effect enables highly stable deep-blue perovskite light-emitting diodes. Angew Chem Int Ed 2024; 63: e202412915.
[48] Wang F, Liu T, Liu Y, et al. A polymeric hole transporter with dual-interfacial interactions enables 25%-efficiency blade-coated perovskite solar cells. Adv Mater 2024; 36: 2412059.
[49] Wang S, Yao H, Zhu W, et al. Stabilization of perovskite lattice and suppression of $Sn^{2+}/Sn^{4+}$ oxidation via formamidine acetate for high efficiency tin perovskite solar cells. Adv Funct Mater 2023; 33: 2215041.
[50] Kang Z, Feng P, Wang K, et al. Synchronous dimension-crystallization engineering enables highly efficient 2D/3D tin perovskite solar cells. Energy Environ Sci 2025; 18: 4108-4119.
[51] Park J, Kim S, Kim W, et al. Stepwise volatilization induced by nature-sourced volatile solid additives improving the efficiency and stability of perovskite solar cells. Energy Environ Sci 2025; 18: 6260-6272.
[52] Su H, Zhang J, Hu Y, et al. Modulation on electrostatic potential of passivator for highly efficient and stable perovskite solar cells. Adv Funct Mater 2023; 33: 2213123.
[53] Wang K, Yu B, Lin C, et al. Constructing stable perovskite with small molecule bridge interface passivation. Adv Energy Mater 2025; 15: 2405571.

[54] Ma M, Jiang X, Zang Z, et al. Suppressing fluoride segregation for high efficiency tin perovskite solar cells. Adv Funct Mater 2024; 34: 2407095.
[55] Wang L, Bi H, Liu J, et al. Exceeding 15% performance with energy level tuning in tin-based perovskite solar cells. ACS Energy Lett 2024; 9: 6238-6244.
[56] Li Z, Jia C, Wu H, et al. In-situ cross-linked polymers for enhanced thermal cycling stability in flexible perovskite solar cells. Angew Chem Int Ed 2025; 64: e202421063.
[57] Xu Y, Wang Q, Lin Z, et al. Synergistic chemical and physical encapsulation strategies enable highly stable and lead leakage-suppressed perovskite solar cells. Interdiscip Mater 2025; 4: 599-609.
[58] Duan W, Chen K, Pi B, et al. A spatial structure regulation strategy modulated the solubility and compactness of novel face-on oriented bisphosphonate-anchored SAMs for efficient inverted perovskite solar cells. Energy Environ Sci 2025; 18: 7231-7244.
[59] Chen L, Fu S, Li Y, et al. On the durability of tin-containing perovskite solar cells. Adv Sci 2024; 11: 2304811.
[60] Macdonald TJ, Lanzetta L, Liang X, et al. Engineering stable lead-free tin halide perovskite solar cells: lessons from materials chemistry. Adv Mater 2023; 35: 2206684.